# Generation of mode-2 internal waves in a two-dimensional stratification by a mode-1 internal wave

Jianjun Liang[1,2*], Tao Du[3], Xiaoming Li[1,4], Mingxia He[2]

[1] Key Laboratory of Digital Earth Science, Institute of Remote Sensing and Digital Earth, Chinese Academy of Sciences, Beijing, China

[2] Ocean Remote Sensing Institute, Ocean University of China, Qingdao, China

[3] College of Oceanic and Atmospheric Sciences, Ocean University of China, Qingdao, China

[4] Key Laboratory of Earth Observation, Hainan Province, Sanya, China

[*] Corresponding Author. Email: liangjj@radi.ac.cn ; Postal address: Institute of Remote Sensing and Digital Earth, Chinese Academy of Sciences, No.9, DengZhuang Nan Road, Haidian, 100094, Beijing, China

**Abstract:** The generation of mode-2 nonlinear internal waves (IWs) by the evolution of a mode-1 IW in a two-dimensional stratification is investigated. A generation model accounting for intermodal interaction is derived based on a multi-modal approach in a weakly nonlinear and non-hydrostatic configuration. The generation model is numerically solved to simulate the evolution of mode-1 and mode-2 IWs in an inhomogeneous pycnocline. The numerical experiments confirm that a mode-2 IW is generated due to linear and nonlinear intermodal interaction. The mode-2 IW continues growing and gradually separates with the mode-1 IW during the generation process. A non-dimensional quantity quantifying the mode-2 IWs' energy is used to investigate the favorable conditions for the formation of mode-2 IWs. The numerical results suggest that the pycnocline strength or depth prominently affects the formation of mode-2 IWs, followed by pycnocline thickness. A weakening or shoaling pycnocline favors the formation of mode-2 IWs by evidently enhancing linear and nonlinear intermodal interaction, whereas a thinning pycnocline favors the process mainly by enhancing nonlinear intermodal interaction. Shortening the front length inhibits nonlinear intermodal interaction while equivalently strengthens the linear intermodal interaction. Increasing the initial mode-1 IW amplitude can noticeably increase the produced mode-2 IW amplitude.

## 1. Introduction

Nonlinear internal waves (IWs) are a common phenomenon in the coastal oceans and marginal seas [1, 2]. Previous studies show that they have profound impacts on a variety of issues, such as offshore drilling operations [3], underwater acoustic propagation [4] and sediment resuspension [5]. In theory, IWs can be described in terms of vertical mode [6]. The first mode (mode-1) IWs have an in-phase displacement of isopycnals in the vertical direction whereas the second mode (mode-2) IWs have an out-phase behavior. Like mode-1 IWs, mode-2 IWs also take two types of waveforms: convex and concave [7]. For mode-2 convex IWs, the upper isopycnals are displaced upward while the lower isopycnals are displaced downward such that a bulge forms in the middle of a water column. Mode-2 IWs mentioned in the following content are all in the convex type. To date, mode-1 IWs have been extensively studied, covering from generation to dissipation, whereas mode-2 IWs have received less attention. Despite of efforts from laboratory experiments, numerical simulations, *in-situ* measurements and satellite observations having been made to investigate the generation of mode-2 IWs, our understanding of their generation within the ocean interior is still incomplete.

The earliest laboratory experiment of mode-2 IWs comes from the work of Davis and Acrivos [8]. They noticed that mode-2 IWs were easily generated by creating a disturbance to a thin density gradient layer, a transition layer between two deep homogeneous layers. Later, Kao and Pao [9] excited a mode-2 IW by inducing the collapse of a mixed region in a thermocline region; Mehta et al. [10] excited a mode-2 IW by allowing the whole head of a gravity current to intrude into a three-layer fluid with a sufficiently wide middle layer. With a focus on a slope-shelf topography,

Helfrich and Melville [11] found that the breaking instability of a mode-1 IW near the shelf break led to the generation of a mode-2 IW. Inspired by the local generation mechanism of IWs in the central Bay of Biscay proposed by New and Pingree [12, 13], Mercier et al.[14] reproduced the process and detected the response of mode-2 IWs.

Numerical experiments have revealed various generation mechanisms for mode-2 IWs. These mechanisms include: (i) a steady flow passing over isolated topography when resonant generation occurs [15]; (ii) a mode-1 IW interacting with a steep sill [16] or shoaling over shelf-slope topography [17, 18]; (iii) impingement of an internal tidal beam on a pycnocline from below when the horizontal phase speed of the tidal beam matches the eigen-speed of mode-2 IWs [19]; (iv) nonlinear disintegration of mode-2 internal tides [20]; (v) polarity conversion of a concave mode-2 IW [21, 22]. Moreover, recent simulation incorporating interaction of barotropic tides with a subcritical ridge shows that the third mechanism works effectively when both the tidal Froude number and contribution to an internal tidal beam from mode-2 waves are high enough [23].

*In-situ* measurements of mode-2 IWs have been reported in the Middle Atlantic Bight [5], on the shelf of the northern South China Sea[7, 24, 25] , on the northern Heng-Chun Ridge south of Taiwan [26], on the New Jersey Shelf [27], and on the Mascarene Plateau [28, 29]. Differing from the former generation mechanisms, Ramp et al. [26] proposed a lee wave mechanism when a tidal current flows over a ridge and Liu et al. [25] suggested a mode-1 IW disintegration mechanism when the mode-1 IW evolves in a horizontally and vertically varying stratification.

Satellite observations of mode-2 IWs are limited, e.g., reported in da Silva et al. [20], Liu et al. [25] and Dong et al.[30]. The study by Dong et al. [30] suggests a different generation mechanism that mode-2 IWs could be induced by an anticyclonic eddy.

Among the various generation mechanisms for mode-2 IWs, the present study aims to clarify the generation of mode-2 IWs by an evolutionary mode-1 IW in a two-dimensional stratification and find the environmental conditions that favor the formation of mode-2 IWs. The two-dimensional stratification is specified by an inhomogeneous pycnocline with a flat bottom, such as a case reported in an in-situ observation [41]. With the goal of quantifying the examined situation as illustrated in Fig.1, a generation model that accounts for intermodal interaction is derived based on a multi-modal approach. The multi-modal approach has been successfully used in multiple studies of internal waves, such as development of internal solitary waves in various thermocline regimes [31], internal tide generation at the continental shelf [32] and multi-modal evolution of wind-generated long internal waves in a closed basin [33]. All of the previous works are restricted to a one-dimensional stratification, that is, density is only varying in the vertical direction. Here, the multi-modal approach is extended to a two-dimensional stratification, that is, density is varying in both horizontal and vertical directions.

The paper is organized as follows. A theoretical generation model for mode-2 IWs is derived by a multi-modal approach in Section 2. In Section 3, various numerical experiments are set up based on the theoretical generation model. Numerical results are discussed in Section 4 and conclusions are summarized in Section 5.

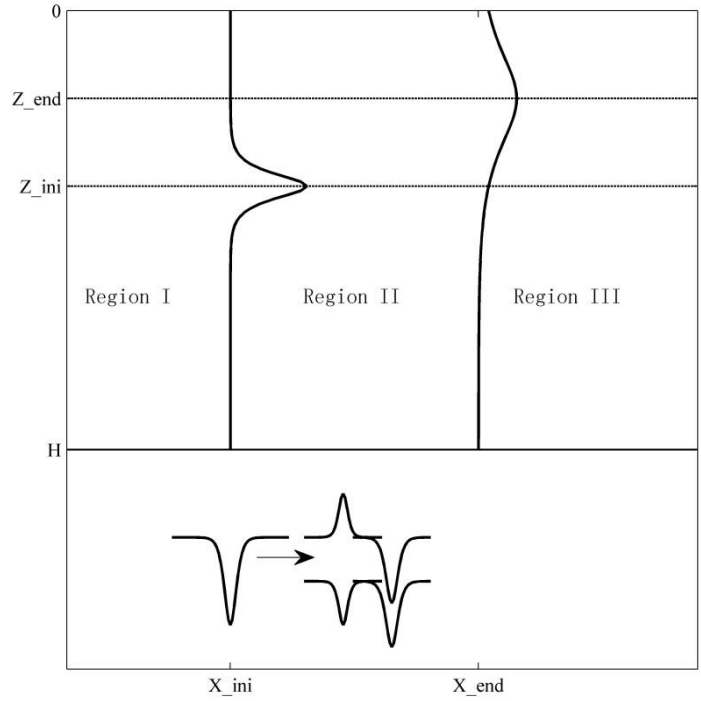

Fig 1. Schematic diagram of the generation of a mode-2 IW by a mode-1 IW evolving in a two-dimensional stratification with an initial one-pycnocline system at $X\_ini$ transforming into another one-pycnocline system at $X\_end$

## 2. Derivation of the generation model

In this section, we examine the generation of mode-2 IWs by a weakly nonlinear non-hydrostatic mode-1 IW evolving in a two-dimensional ambient density field. A geostrophic current, $(0,V,0)$, emerges within the region with horizontal density gradient and vanishes outside the region. The system of equations governing the wave motion consists of the continuity equation, the Euler equations and the incompressibility condition. They take the form in a two-dimensional configuration with the Boussinesq approximation:

$$u_x + w_z = 0,$$
$$u_t + uu_x + wu_z - fv = -p_x,$$
$$v_t + uv_x + wv_z + uV_x + wV_z + fu = 0, \quad (1)$$
$$w_t + uw_x + ww_z = -p_z - \sigma,$$
$$\sigma_t + u\sigma_x + w\sigma_z = N^2 w + M^2 u.$$

Here, subscripts denote partial derivatives; the surface of the fluid is located at $z = 0$ and $z$ is positive upward; $u$, $v$ and $w$ are the longitudinal, transverse and vertical velocity components induced by waves, respectively; $t$ is the time; $f$ is the constant Coriolis parameter; $p$ is the density normalized pressure perturbation; $\sigma$ is the density perturbation with respect to its local static value (multiplied by $g/\rho_0$; $g$ the gravitational acceleration and $\rho_0$ a constant reference density); $N^2 = -g\overline{\rho}_z/\rho_0$ is the Brunt-Väisälä frequency; and $M^2 = -g\overline{\rho}_x/\rho_0$ is the horizontal analogy to $N^2$. The geostrophic current $V$ satisfies the thermal wind relation $V_z = M^2/f$. In the study, we consider a frontal zone in which the background density $\overline{\rho}(x,z)$ has a linear variation in the longitudinal direction. Therefore, the $uV_x$ term vanishes in the following equations because the term involves a second derivative of the background density with respect to the longitudinal coordinate [41].

The boundary conditions of Eq. (1) are

$$w(z=0) = w(z=-H) = 0. \quad (2)$$

In order to investigate weakly nonlinear non-hydrostatic internal waves, it is convenient to scale the equations. We use the following time and space scales:

$$[t] = 1/\omega \ ; \ [x] = L \ ; [z] = H, \quad (3)$$

where $[t]$ and $[x]$ are characteristic values of the period and wave length of the IWs. The evolution of IWs in a two-dimensional stratification is governed by the following non-dimensional parameters:

$$\varepsilon = U/L\omega \; ; \; \delta = H/L \; ; \mu = f_0/\omega \; ; , \tag{4}$$

where $\varepsilon$ is a parameter of nonlinearity, $\delta$ of non-hydrostatic dispersion, $\mu$ of Coriolis dispersion ; $U$ is the characteristic value of the wave-induced horizontal current and $f_0$ is the characteristic value of Coriolis parameter.

The scales of dynamical variables are found by considering the linear version of Eq. (1). Thus, we obtain:

$$[w] = \delta U \; ; [v] = \mu U \; ; \; [p] = U\omega L \; ; [\sigma] = U\omega/\delta \tag{5}$$

With the aforementioned scales, Eq. (1) becomes non-dimensional,

$$\begin{aligned}
&u_x + w_z = 0, \\
&u_t + \varepsilon(uu_x + wu_z) - \mu^2 fv = -p_x, \\
&v_t + \varepsilon(uv_x + wv_z) + \delta/f_0^2 \, wM^2/f + fu = 0, \\
&w_t + \varepsilon(uw_x + ww_z) = -1/\delta^2 \, p_z - 1/\delta^2 \, \sigma, \\
&\sigma_t + \varepsilon(u\sigma_x + w\sigma_z) = N^2 w + \delta\mu^2/f_0^2 \, M^2 u,
\end{aligned} \tag{6}$$

where we use the same symbols, for convenience, to denote dimensionless quantities expect for the horizontal stratification $M^2$. We do not scale $M^2$ in Eq. (6) because its characteristic value is not clear as yet. The scaling of $M^2$ is based on the assumption that the horizontal stratification varies slowly with respect to the length scale of IWs. Thus, we introduce a "slow" horizontal coordinate $\xi = \gamma x$ with

$$\gamma = \frac{IW \; length \; scale}{Horizontal \; stratification \; length \; scale} \ll 1 \tag{7}$$

In particular, only local static density is the function of $\xi$ and other fields depend on both $\xi$ and $x$. Introducing the coordinate $\xi$ into Eq. (6) yields

$$u_x + \gamma u_\xi + w_z = 0,$$
$$u_t + \varepsilon(uu_x + wu_z) - \mu^2 fv = -p_x - \gamma p_\xi,$$
$$v_t + \varepsilon(uv_x + wv_z) + \delta v w M^2/f + fu = 0, \quad (8)$$
$$w_t + \varepsilon(uw_x + ww_z) = -1/\delta^2 \, p_z - 1/\delta^2 \, \sigma,$$
$$\sigma_t + \varepsilon(u\sigma_x + w\sigma_z) = N^2 w + \delta\mu^2 v M^2 u,$$

where the non-dimensional parameter $v$ is expressed as $v = M_0^2/f_0^2$ with $M_0^2$ being the characteristic value of horizontal stratification. Meanwhile, we can systematically examine the orders of magnitude of the various terms in Eq. (8) according to the KdV theory for weakly nonlinear IWs and their typical scales in oceanic conditions. The related quantities are scaled as follows:

$$U = O(10^{-2}\,\mathrm{ms^{-1}}); L = O(10^3\,\mathrm{m}); \omega = O(10^{-3}\,\mathrm{s^{-1}}); f_0 = O(10^{-4}\,\mathrm{s^{-1}});$$
$$H = O(10^2\,\mathrm{m}); \varepsilon = O(10^{-2}); \delta = O(10^{-1}); \mu = O(10^{-1}). \quad (9)$$

Substituting Eq. (9) into the fifth equation in Eq.(8), we yield $M_o^2 = O(10^{-7}\,\mathrm{s^{-2}})$ in order that the horizontal stratification itself induced modal coupling appears in the same order approximation as those induced by nonlinear advection terms. Moreover, the parameter $\gamma$ needs to satisfy $\gamma = O(10^{-2})$ to ensure the horizontal stratification itself induced $u_\xi$ in the first equation in Eq. (8) consistently appear in the second order expansion.

In terms of the above scaling analysis, we can obtain relationships between terms of like order in $\varepsilon$. Then, it is feasible to expand the dynamical variables in the order of $\varepsilon$:

$$(u, v, w, p, \sigma) = \left(u^{(0)}, v^{(0)}, w^{(0)}, p^{(0)}, \sigma^{(0)}\right) + \varepsilon\left(u^{(1)}, v^{(1)}, w^{(1)}, p^{(1)}, \sigma^{(1)}\right) + \cdots \quad (10)$$

Substituting Eq. (10) into Eq. (8), we arrive at the leading-order linear equation set:

$$\begin{aligned}
u_x^{(0)} + w_z^{(0)} &= 0, \\
u_t^{(0)} + p_x^{(0)} &= 0, \\
v_t^{(0)} + w^{(0)}V_z + fu^{(0)} &= 0, \\
p_z^{(0)} + \sigma^{(0)} &= 0, \\
\sigma_t^{(0)} - N^2 w^{(0)} &= 0.
\end{aligned} \qquad (11)$$

From Eq. (11), the linear equation for $w^{(0)}$ is obtained:

$$w_{zztt}^{(0)} + N^2 w_{xx}^{(0)} = 0 \qquad (12)$$

As the stratification varies less slowly than IWs in the longitudinal direction, the solution of $w^{(0)}$ can be found by separation of variables:

$$w^{(0)} = \sum_n W_n^{(0)}(x,\xi,t) \phi_n(\xi,z), \qquad (13)$$

where $\phi_n(\xi,z)$ is the eigenfunction of a wave mode, determined by the solution of the eigenvalue problem along the vertical line for every $\xi$:

$$\frac{\partial^2 \phi_n}{\partial z^2} + \frac{N^2(\xi,z)}{c_n^2(\xi)} \phi_n = 0; \qquad \phi_n(z=0) = \phi_n(z=-H) = 0, n = 1,2,3,\cdots, \qquad (14)$$

where $c_n$ is the linear phase speed. The corresponding orthogonality relation is

$$\int_{-H}^{0} \phi_n' \phi_m' dz = \frac{I_n}{c_n^2} \delta_{mn}; \qquad \int_{-H}^{0} N^2 \phi_n \phi_m dz = I_n; \\
I_n = \int_{-H}^{0} N^2 \phi_n^2 dz, \qquad (15)$$

where $\delta_{mn}$ is the Kroneker's delta.

All dependent variables are now expanded using the consistency implied by Eq. (11):

$$\begin{aligned}
u^{(0)} &= \sum_n U_n^{(0)}(x,\xi,t) \phi_n'(\xi,z), \\
p^{(0)} &= \sum_n P_n^{(0)}(x,\xi,t) \phi_n'(\xi,z), \\
\sigma^{(0)} &= \sum_n Z_n^{(0)}(x,\xi,t) N^2(\xi,z) \phi_n(\xi,z),
\end{aligned} \qquad (16)$$

The variable $v^{(0)}$ is not expanded because it is a linear combination of $w^{(0)}$ and $u^{(0)}$. Moreover, as shown below, only $v^{(0)}$ works in the final governing equation.

Substituting Eq. (16) into Eq. (11), employing Eq. (15), and eliminating $W_n^{(0)}$ and $P_n^{(0)}$, we obtain the leading-order coupled pair of evolution equations

$$U_{nx}^{(0)} + Z_{nt}^{(0)} = 0,$$
$$U_{nt}^{(0)} + c_n^2 Z_{nx}^{(0)} = 0. \tag{17}$$

Continuing to perform asymptotic analysis to the next order, we obtain the following inhomogeneous equation set:

$$u_x^{(1)} + w_z^{(1)} = -u_\xi^{(0)},$$
$$u_t^{(1)} + p_x^{(1)} = -p_\xi^{(0)} - \left\{ u^{(0)} u_x^{(0)} + w^{(0)} u_z^{(0)} \right\} + f v^{(0)},$$
$$p_z^{(1)} + \sigma^{(1)} = -w_t^{(0)},$$
$$\sigma_t^{(1)} - N^2 w^{(1)} = -\left\{ \left( u^{(0)} \sigma^{(0)} \right)_x + \left( w^{(0)} \sigma^{(0)} \right)_z \right\} + M^2 u^{(0)}. \tag{18}$$

In Eq. (18), the term $u_\xi^{(0)}$ in the first equation and $M^2 u^{(0)}$ in the last equation are the leading-order effect of slowing varying horizontal stratification; the bracket terms in the second equation contain the leading-order nonlinear acceleration and the term $f v^{(0)}$ contains leading-order geostrophic current correction; the term $w_t^{(0)}$ in the third equation is leading-order non-hydrostatic correction; and the bracket terms in the last equation define the leading-order buoyancy flux correction.

We expand the second-order variables $u^{(1)}$, $w^{(1)}$, $p^{(1)}$, $w^{(1)}$, and $\sigma^{(1)}$ in the same manner as in Eq. (16). After some algebraic manipulation, the coupled pair of evolution equations at the next order approximation are obtained in the form:

$$U_{nt}^{(1)} + c_n^2 Z_{nx}^{(1)} = -\sum_{ij} \left\{ a_{nij}^{(u)} U_i^{(0)} U_{jx}^{(0)} + b_{nij}^{(u)} U_{ix}^{(0)} U_j^{(0)} \right\} + \sum_i d_{ni} U_{ixxt}^{(0)}$$
$$- \frac{\partial c_n^2}{\partial \xi} Z_n^{(0)} - c_n^2 Z_{n\xi}^{(0)} - \sum_i \left\langle \phi_n' \frac{\partial \phi_i'}{\partial \xi} \right\rangle c_i^2 Z_i^{(0)} + \left\langle f v^{(0)} \phi_n' \right\rangle \tag{19}$$

$$Z_{nt}^{(1)} + U_{nx}^{(1)} = -\sum_{ij}\left\{a_{nij}^{(\sigma)}\left(U_i^{(0)}Z_j^{(0)}\right)_x + b_{nij}^{(\sigma)}U_{ix}^{(0)}Z_j^{(0)}\right\} - U_{n\xi}^{(1)}$$
$$-\sum_i\left\langle\phi_n'\frac{\partial\phi_i'}{\partial\xi}\right\rangle U_i^{(0)} + \sum_i\frac{U_i^{(0)}}{I_n}\int_H^0\phi_i'\phi_n M^2 dz \quad (20)$$

The coefficients appearing in Eqs. (19, 20) are defined by the following set of relations:

$$a_{nij}^{(u)} = \left\langle\phi_n'\phi_i'\phi_j'\right\rangle, b_{nij}^{(u)} = \left\langle\frac{N^2}{c_j^2}\phi_n'\phi_i\phi_j\right\rangle,$$
$$a_{nij}^{(\sigma)} = \left\langle\frac{N^2}{c_n^2}\phi_n\phi_i'\phi_j\right\rangle, b_{nij}^{(\sigma)} = \left\langle\frac{N^2}{c_n^2}\phi_n'\phi_i\phi_j\right\rangle, \quad (21)$$
$$d_{ni} = \left\langle\phi_n\phi_i\right\rangle,$$

where $\langle\cdots\rangle \equiv \dfrac{c_n^2}{I_n}\int_{-H}^0(\cdots)dz.$

Combing Eq.(17) and Eqs. (19, 20), and transforming the coordinate system back to ordinary physical coordinates lead to the final weakly-nonlinear evolution set:

$$U_{nt} + \left(c_n^2 Z_n\right)_x = -\sum_{ij}\left\{a_{nij}^{(u)}U_iU_{jx} + b_{nij}^{(u)}U_{ix}U_j\right\}$$
$$+\sum_i d_{ni}U_{ixxt} - \sum_i r_{ni}c_i^2 Z_i + \left\langle fv\phi_n'\right\rangle \quad (22)$$

and

$$Z_{nt} + U_{nx} = -\sum_{ij}\left\{a_{nij}^{(\sigma)}\left(U_iZ_j\right)_x + b_{nij}^{(\sigma)}U_{ix}Z_j\right\}$$
$$-\sum_i s_{ni}U_i, \quad (23)$$

where

$$r_{ni} = \left\langle\phi_n'\frac{\partial\phi_i'}{\partial x}\right\rangle; s_{ni} = r_{ni} - \frac{1}{I_n}\int_{-H}^0\phi_n\phi_i'M^2 dz. \quad (24)$$

Equations (22, 23) comprise four kinds of modal coupling terms: nonlinear ones involving terms with $a$ and $b$, non-hydrostatic ones involving terms with $d$, horizontal stratification induced ones involving terms with $S$ and $r$, and geostrophic current induced ones involving terms with $fv$.

In the present study, we restrict the analysis to the first and second vertical modes because they are energetically dominant in many cases. Thus, the evolution equation for mode-1 IWs is

$$U_{1t} + \left(c_1^2 Z_1\right)_x = -\sum_{i=1}^{2}\sum_{j=1}^{2}\left(a_{1ij}^u U_i U_j + b_{1ij}^u U_{ix} U_j\right)$$
$$+ \sum_{i=1}^{2} d_{1i} U_{ixxt} - \sum_{i=1}^{2} r_{1i} c_i^2 Z_i + \left\langle fv\phi_1'\right\rangle,$$
$$Z_{1t} + U_{1x} = -\sum_{i=1}^{2}\sum_{j=1}^{2}\left(a_{1ij}^\sigma \left(U_i Z_j\right)_x + b_{1ij}^\sigma U_{1x} Z_j\right)$$
$$- \sum_{i=1}^{2} s_{1i} U_i,$$

(25)

and for mode-2 IWs is

$$U_{2t} + \left(c_2^2 Z_2\right)_x = -\sum_{i=1}^{2}\sum_{j=1}^{2}\left(a_{2ij}^u U_i U_j + b_{2ij}^u U_{ix} U_j\right)$$
$$+ \sum_{i=1}^{2} d_{2i} U_{ixxt} - \sum_{i=1}^{2} r_{2i} c_i^2 Z_i + \left\langle fv\phi_2'\right\rangle,$$
$$Z_{2t} + U_{2x} = -\sum_{i=1}^{2}\sum_{j=1}^{2}\left(a_{2ij}^\sigma \left(U_i Z_j\right)_x + b_{2ij}^\sigma U_{2x} Z_j\right)$$
$$- \sum_{i=1}^{2} s_{2i} U_i.$$

(26)

The transverse velocities $v$ in Eqs. (25, 26) are solved from $v^{(0)}$ in Eq. (11). Equations (25, 26) are solved numerically using finite-difference method: 4th-order compact scheme in space [34, 35] and 3rd-order Adams-Bashforth scheme in time [36]. Due to nonlinear inter-modal interaction, the model result easily blows up during wave evolution process, so we use a spatial low-pass filter every 5 time steps to prevent instabilities on the grid scale [37]. In the following numerical simulation, the grid resolution is $dx = 50$m and time step is $dt = 5$s. The choice of this configuration ensures sufficiently resolving the process that we concern. In addition, a linear simulation with $M^2 = 0$ shows the numerical scheme was robust.

## 3. Set-up of the generation model

In this study, we assume a flat bottom with a constant depth $H = 200\,\text{m}$. The two-dimensional background density is constructed in accordance with Vlasenko et al. [41].

First, the horizontally uniform density fields in Region I and Region III (Fig.1) are prescribed. Then, the density field in Region II, a front zone, is obtained via a linear interpolation of the density difference between Region I and Region III. The background density field in Region I and Region III is represented by a hyperbolic tangent function [20], respectively:

$$\overline{\rho}(x,z) = \rho_0 \exp\left[\frac{\Delta\rho_{ini}}{2\rho_0}\tanh\left(\frac{-2(z-z_{ini})}{d_{ini}}\right)\right], \qquad 0 \leq x \leq X\_\text{ini}, -H \leq z \leq 0,$$
$$\overline{\rho}(x,z) = \rho_0 \exp\left[\frac{\Delta\rho_{end}}{2\rho_0}\tanh\left(\frac{-2(z-z_{end})}{d_{end}}\right)\right], \qquad x \geq X\_\text{end}, -H \leq z \leq 0, \tag{27}$$

where $\Delta\rho_{ini}$, $z_{ini}$ and $d_{ini}$ are the constant density difference, depth and thickness of an initial pycnocline, respectively; $\Delta\rho_{end}$, $z_{end}$ and $d_{end}$ are the constant density difference, depth and thickness of a final pycnocline, respectively. The set-up of these parameters needs efforts to be representative of oceanic conditions.

We refer to the characteristic values of pycnoclines presented in [20] to set the physical parameters in Eq. (27). In Region I, the three parameters $\Delta\rho_{ini}$, $z_{ini}$ and $d_{ini}$ are taken to be $2.7004\,\text{kgm}^{-3}$, $-80\,\text{m}$ and $80\,\text{m}$, respectively, and the resulting maximum buoyancy frequency $N_{I\_\text{max}}$ is $0.018\,\text{s}^{-1}$. The buoyancy frequency profile in Region I keeps constant in the numerical simulations, whereas the buoyancy frequency profile in Region III are varied to represent possible variations of the initial pycnocline by changing the three parameters $\Delta\rho_{end}$, $z_{end}$ and $d_{end}$. Among them, the parameter $\Delta\rho_{end}$ is taken to be 0.8335, 2.7004 and 5.6342 $\text{kgm}^{-3}$ to make the maximum buoyancy frequency $N_{III\_\text{max}}$ equal 0.010, 0.018 and $0.026\,\text{s}^{-1}$, respectively. The pycnocline depth $z_{end}$ is taken to be -20, -80 and $-100\,\text{m}$, respectively, and the pycnocline thickness $d_{end}$ is taken to be 40, 80 and 120 m, respectively.

With the definition in Eq. (27), the length of Region II is $L=X\_end - X\_ini$. The set-up of its value is dependent on the requirement that the horizontal stratification induced modal interaction should be at the same order as the nonlinear advection induced ones, so a typical length of Region II is set to 20 km and another two values, 5 and 35 km are also taken into account for sensitivity runs.

The initial incident IW at $x = X\_ini$ is given as the mode-1 solitary wave solution of the KdV equation [6]:

$$Z_1(x,t) = A \sec h^2(\frac{x - X\_ini - C_{iw}t}{L_w}),$$
$$C_{iw} = c_1(x = X\_ini) + \frac{1}{3}\alpha A \quad (28)$$

where A is the wave amplitude, $C_{iw}$ is the nonlinear phase speed, $L_w$ is the half width, and $\alpha$ is the nonlinear parameter. The parameters $C_1$ and $\alpha$ can be obtained by substituting the known stratification profile at $x = X\_ini$ into Eq. (14) [6]. The wave amplitude and half width are set to -2 and 1040 m to meet the scaling conditions in Eq.(9). Besides, we take $A = -5.5$m and $A = -10.5$m to test the influence of wave amplitude.

Based on the above considerations, eleven cases were carried out and their parameters are listed in Table 1.

## 4. The experiment results and discussions

### 4.1 Typical generation process

This section is devoted to elaborating the generation process of a mode-2 IW excited by a clean mode-1 IW propagating into a two-dimensional stratification, characterized by an inhomogeneous pycnocline. As the essential features of the generation are the same in the eleven cases listed in Table 1, we use the numerical results of E4 with a thickening pycnocline to demonstrate the problem.

The wave generation is illustrated by plotting the vertical displacement field of IWs given by $\eta_n(x,z,t) = Z_n(x,t)\phi_n(z), n=1,2$. Their distribution at $t=0.25\text{h}$ is shown in Fig.2. It can be seen that Fig. 2(a) contains a mode-1 IW with the positions of fluid particles consistently below their equilibrium levels in the vertical direction whereas an opposite manifestation indicating a mode-2 IW is shown in Fig.2 (b). At the moment, the mode-2 IW almost coincides with the mode-1 IW in the horizontal space, as indicated by the dash line. The appearance indicates that the mode-1 IW just brings about the generation of the mode-2 IW. Essentially, the generation event arises from the intermodal interaction incorporated in the generation model. The model comprises two kinds of intermodal interaction. One kind is linear, relating to horizontal stratification and its associated geostrophic current, while the second kind is related to nonlinear advection and non-hydrostatic effect. The significance of the two kinds of intermodal interaction in favoring the generation of mode-2 IWs will be presented in subsection 4.2.

As time goes on, the mode-2 IW grows and gradually separates with the mode-1 IW, suggested by a comparison between Fig. 2 and Fig.3. In Fig.3, the mode-2 IW lags behind the mode-1 IW by about 500 m at $t=1\text{h}$. From t=0.25 h to t=1.0 h, the mode-1 IW propagates from x=15.85 km (Fig.2) to x=18.25 km (Fig.3). Thus, the propagation speed of the mode-1 IW is $C_1 = \frac{(18.25-15.75)\times 1000}{(1.0-0.25)\times 3600} \approx 0.89 \text{ms}^{-1}$. On the other hand, the mean linear phase speed of the mode-1 IW in the spatial range is approximately $0.90 \text{ ms}^{-1}$. So, the numerical and theoretical phase speeds coincide with each other for the mode-1 IW. However, within the same time range, the numerical phase speed for mode-2 IW is much larger than the linear phase speed. The reason may be due to the nonlinear coupling of mode-1 and mode-2 IWs. Meanwhile, with the aim of clarifying the development of the mode-2 IW during the generation process, the temporal variation of mode-2 wave amplitude $Z_2$ is shown in Fig. 4. As expected, the mode-2 wave amplitude continuously increases and reaches a maximum at about $t=0.75\text{h}$. Afterwards, the wave amplitude varies much slowly around the maximum.

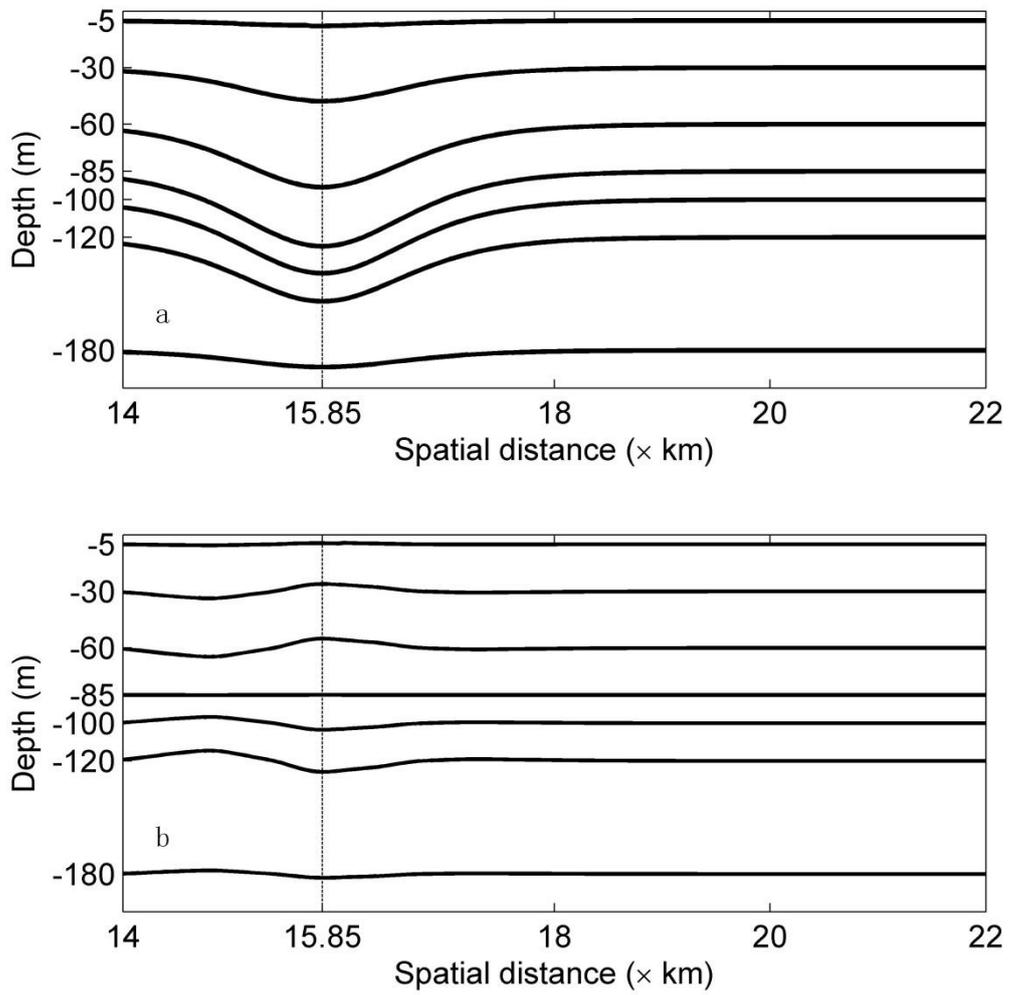

Fig 2. The vertical displacement filed of a mode-1 IW (a) and a mode-2 IW (b) at $t = 0.25$h for

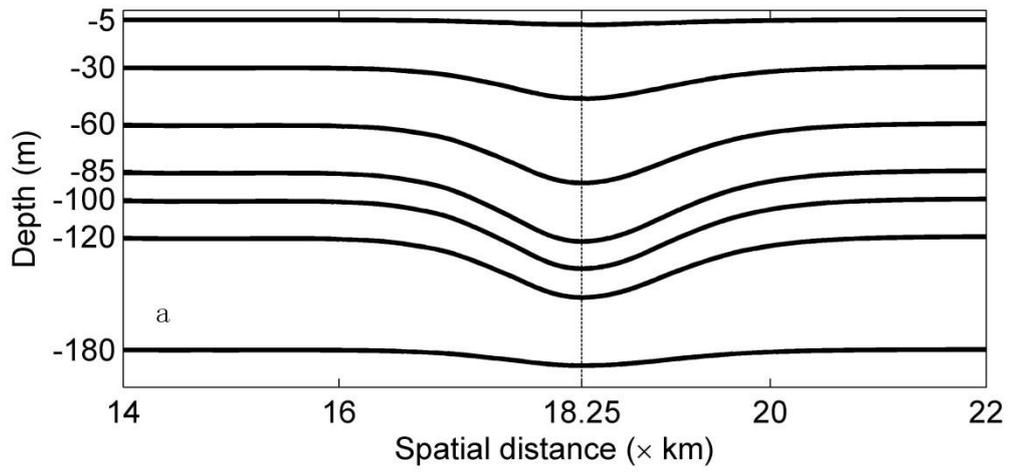

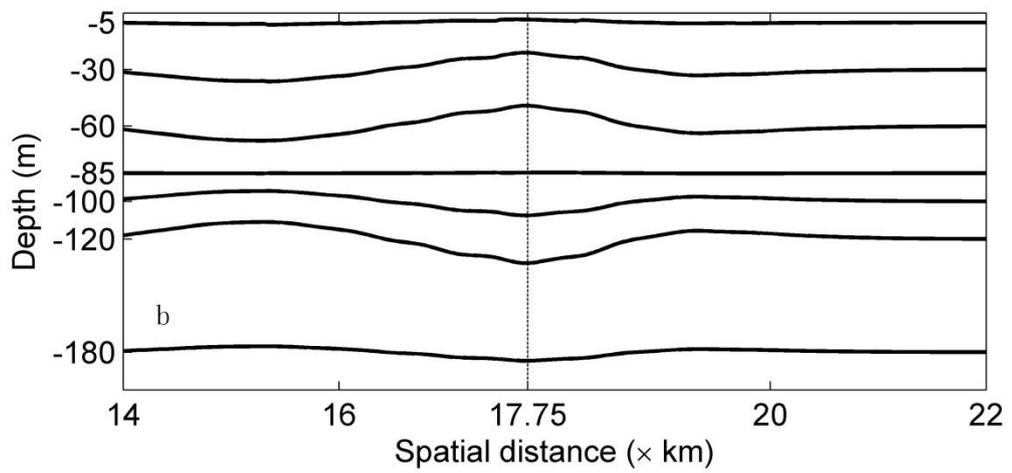

Fig 3. The vertical displacement filed of a mode-1 IW (a) and a mode-2 IW (b) at $t = 1.0\text{h}$ for

E4

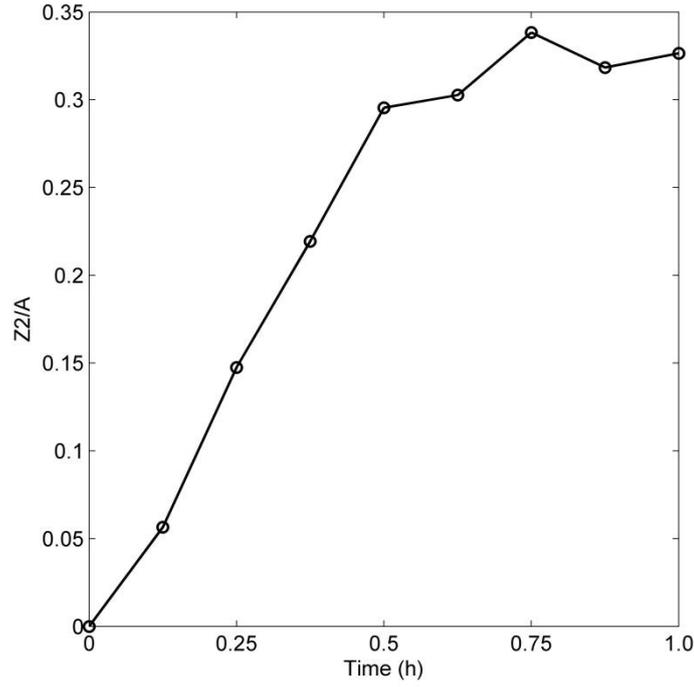

Fig 4. The variation of mode-2 wave amplitude $Z_2$ normalized by the initial mode-1 wave amplitude $A$ versus time for E4

**4.2 Favorable environmental condition for model -2 IWs' formation**

In the section, we focus on identifying which environmental conditions favor the formation of mode-2 IWs by the evolution of an initial mode-1 IW in a two-dimensional stratification. The favorable condition is defined as one that induces the mode-2 wave energy to account for the largest portion of an initial mode-1 wave energy. Therefore, a non-dimensional quantity $R_2 = \dfrac{E_2}{E_1(t=0)}$ is put forward to quantify the environmental condition, where $E_2$ is the modal energy of mode-2 IWs and $E_1(t=0)$ is the modal energy of an initial mode-1 IW. The modal energies are defined in the same way as Sakai and Redekopp [33]:

$$E_1(t=0) = \int_0^{L_c} \left\{ \frac{1}{2}\frac{I_1}{c_1^2}\left[U_1^2 + d_{11}U_{1x}^2\right] + \frac{1}{2}I_1\left[Z_1^2 - a_{111}^{(\sigma)}Z_1^3\right]\right\}dx,$$

$$E_2 = \int_0^{L_c} \left\{ \begin{array}{l} \frac{1}{2}\frac{I_2}{c_2^2}\left[U_2^2 + d_{22}U_{2x}^2 + d_{21}U_{1x}U_{2x}\right] + \frac{1}{2}I_2\left[Z_2^2 - a_{222}^{(\sigma)}Z_2^3\right] \\ -\frac{1}{2}I_1 Z_2^2 Z_1\left[2a_{122}^{(\sigma)} + b_{122}^{(\sigma)}\right] + \frac{1}{2}\int_{-H}^0 V_2^2 dz + \frac{1}{2}\int_{-H}^0 V_1 V_2 dz \end{array} \right\} dx, \quad (29)$$

where $L_c$ is the length of the computational domain, $V_1 = \int_{-H}^0 N^2 \phi_1 v dz / I_1$, $V_2 = \int_{-H}^0 N^2 \phi_2 v dz / I_2$, and they vanish in $E_1(t=0)$ because they are taken to be zero at the initial time of numerical simulations. Here, a point needs to be addressed. A little difference between Eq. (29) and the derived definition of Sakai and Redekopp [33] exists, which lies in the additional consideration of transverse velocity in Eq. (29) because it is nearly at the same order as the vertical velocity according to the scaling conditions in the present study.

### 4.2.1 Overall description of environmental effects

Figure 5 shows the variation of $R_2$ versus time under different environmental conditions, including the variations of pycnocline depth (Fig.5a), pycnocline thickness (Fig.5b), pycnocline strength (Fig.5c) and front length (Fig.5d). In most cases, $R_2$ initially increases with time and saturates at $t = 2.5\text{h}$, whereas a persistent increase of $R_2$ occurs in E1 and E5. Although $R_2$ does not reach a steady state in both cases, they evidently overwhelm the other cases in the sensitive experiments for the pycnocline depth and pycnocline strength, respectively, which suffices for investigating the favorable conditions for mode-2 IWs' formation.

An overall observation of Fig. 5 reveals that the pycnocline depth, pycnocline thickness, pycnocline strength and front length change $R_2$ in varying degrees. The large change of $R_2$ occurs in the control tests of pycnocline strength or depth, reaching nearly 10%. The secondary change occurs for pycnocline thickness, reaching nearly 5%. The smallest change is caused by front length, decreasing to only 1%.

From the above analysis, we can conclude that the formation of mode-2 IWs is most sensitive to the pycnocline strength or depth, followed by pycnocline depth, and least sensitive to front length.

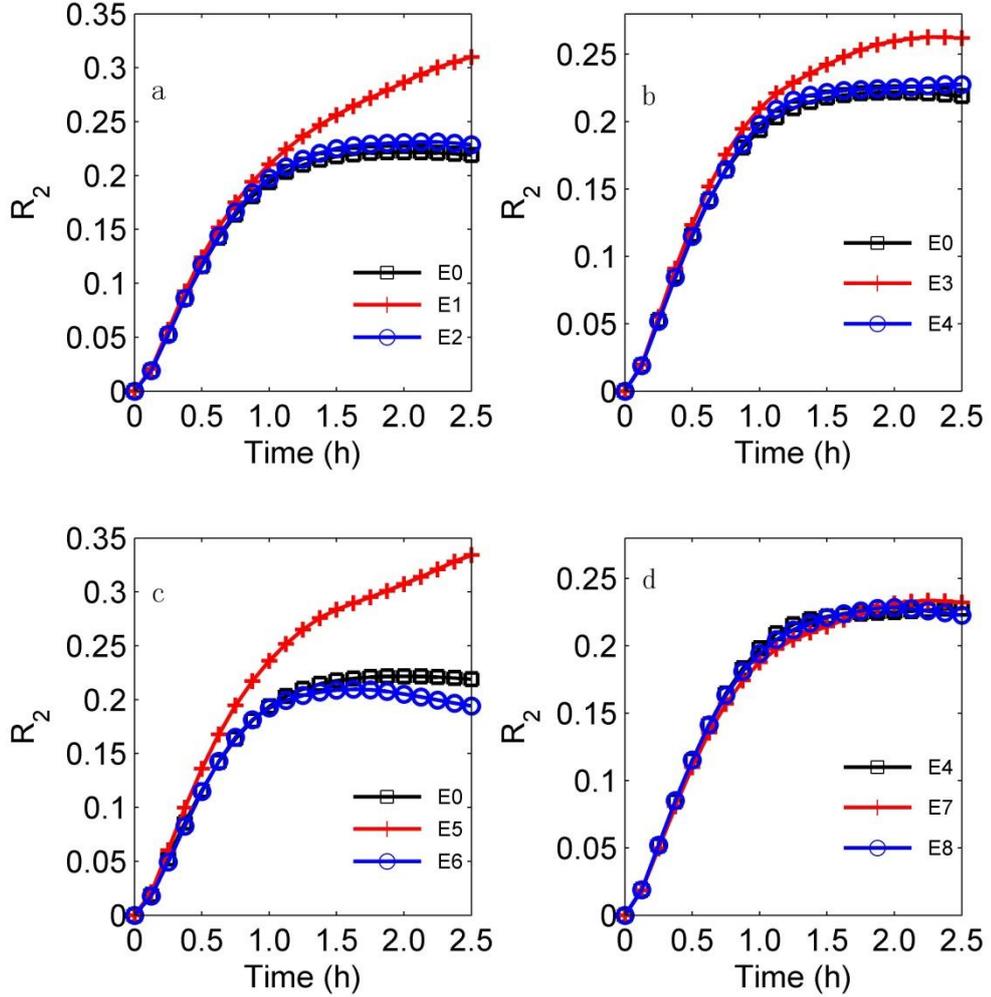

Fig 5. The ratio of mode-2 IWs' energy to an initial mode-1 IW's energy, $R_2$, plotted against time for different pycnocline depths(a), pycnocline thicknesses (b) , pycnocline strengths(c) and front lengths(d)

### 4.2.2 Effect of pycnocline depth

Fig. 5(a) depicts the evolution of $R_2$ for cases E0 (homogeneous pycnocline), E1 (shoaling pycnocline) and E2 (deepening pycnocline). One observes that case E1 stands out in the control test of pycnocline depth. The E1 has a continuously increasing $R_2$ with a value of 31% at $t = 2.5\text{h}$, while the E2 and E3 have a saturated

$R_2$ with a maximum of about 22% at $t = 2.5\text{h}$. The results indicate that a shoaling pycnocline favors the formation of mode-2 IWs. The reason lies in that a shoaling pycnocline can evidently enhance the IW growth, which has been revealed by the modelling study of Buijsman et al. [42] and again evidenced by the case E1 which features a simultaneous increase of mode-1 and mode-2 wave amplitude versus time.

In more detail, the effects of linear and nonlinear intermodal interaction are analyzed in Fig. 6. Shown is the evolution of $R_2$ obtained by independently considering linear or nonlinear intermodal interaction in the generation model. When considering linear intermodal interaction, the intermodal interaction terms caused by nonlinear advection and non-hydrostatic dispersion are set to zero and vice versa. The results indicate that a shoaling pycnocline favors the formation of mode-2 IWs by simultaneously enhancing the linear and nonlinear intermodal interaction compared to a homogeneous or deepening pycnocline. Furthermore, the nonlinear intermodal interaction overwhelms linear intermodal interaction during the considered time range and plays a dominant role at the initial growth stage of mode-2 IWs. After the initial stage, the nonlinear modal interaction remains steady. The linear intermodal interaction gradually strengthens and contributes to the continuous increase of $R_2$ in Fig. 5(a).

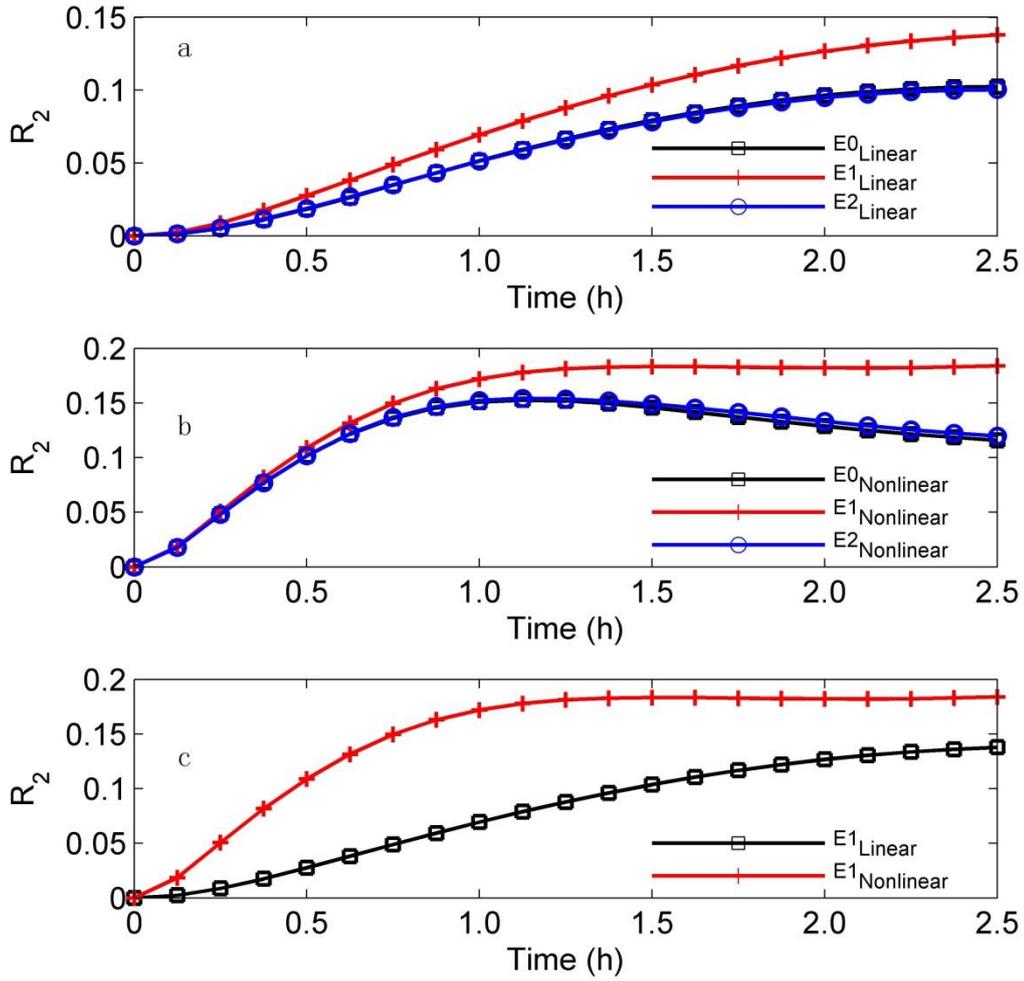

Fig.6 Comparison of $R_2$ for linear tests of E0, E1 and E2 (a), for nonlinear tests of E0, E1 and E2 (b) and for linear and nonlinear tests of E1 (c)

**4.2.2 Effect of pycnocline thickness**

Fig. 5(b) depicts the evolution of $R_2$ for cases E0 (homogeneous pycnocline), E3 (thinning pycnocline) and E4 (thickening pycnocline). The results suggest that a thin pycnocline favors the formation of mode-2 IWs, consistent with the analytical study of [7] that a thin middle layer benefits the survival of mode-2 IWs. In terms of Fig.7, we explain the favoring mechanism by means of a strengthening nonlinear intermodal interaction (Fig.7b) because the effects of linear intermodal interaction are nearly the same as in the three cases (Fig.7a). The strengthening of nonlinear intermodal interaction probably results from wave trapping in a narrowing waveguide, which benefits the energy transfer from mode-1 IWs to mode-2 IWs.

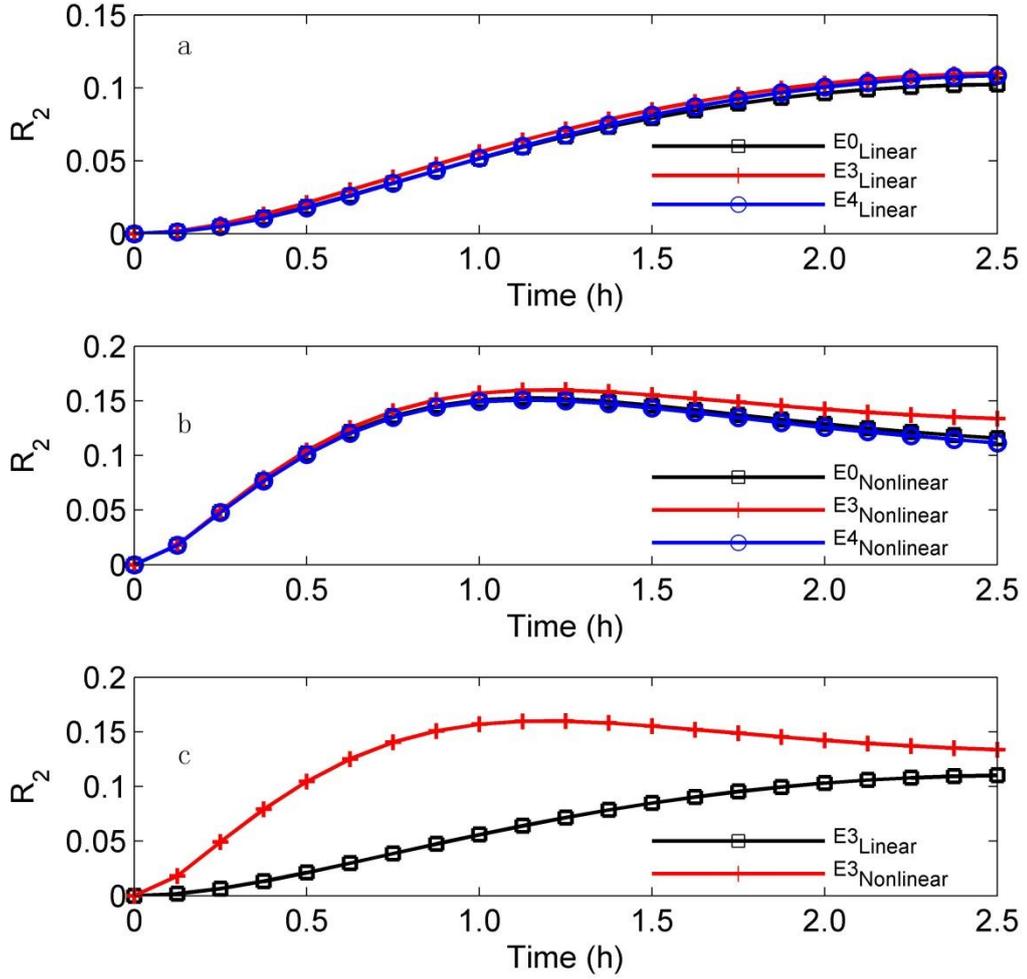

Fig. 7 Comparison of $R_2$ for linear tests of E0, E3 and E4 (a), for nonlinear tests of E0, E3 and E4 (b) and for linear and nonlinear tests of E3 (c)

**4.2.3 Effect of pycnocline strength**

Fig. 5(c) depicts the evolution of $R_2$ for cases E0 (homogeneous pycnocline), E5 (weakening pycnocline) and E6 (strengthening pycnocline). The curves clearly show that a weakening pycnocline greatly favors the formation of mode-2 IWs, probably because a weakly stratified regime suits the intermodal interaction as a result of a weak waveguide effect. The argument coincides with the modelling study of Gerkema [40] that a stronger waveguide effect, a pycnocline, inhibits the mode coupling of a wave beam propagating from a weaker waveguide environment.

Furthermore, by comparing Fig.8 and Fig.6, we conclude that a weakening pycnocline favors the formation of mode-2 IWs by the same mechanism as the shoaling pycnocline. In addition, an interesting point to note about the plot is that a strengthening pycnocline promotes the linear intermodal interaction whereas inhibits nonlinear intermodal interaction compared to a homogeneous pycnocline, thus yielding the nearly same trends for E0 and E6 in Fig.5(c).

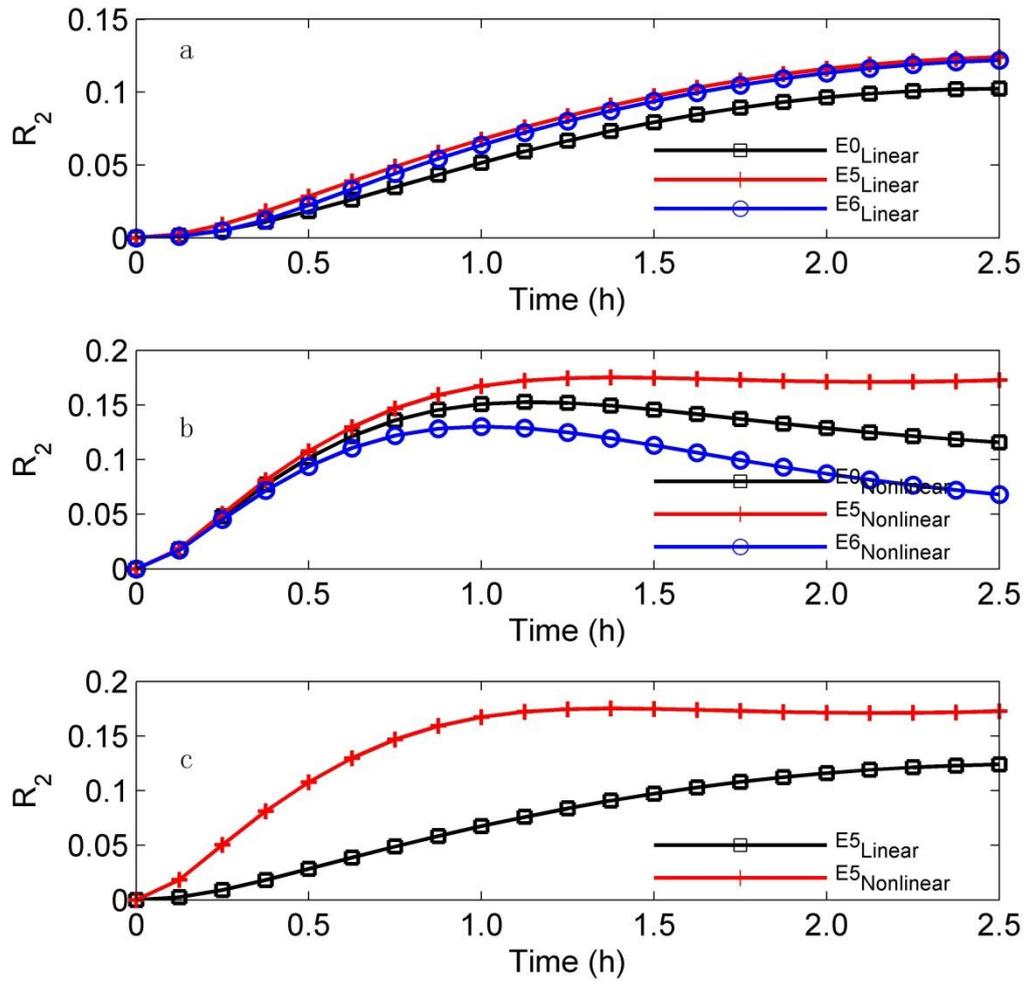

Fig. 8 Comparison of $R_2$ for linear tests of E0, E5 and E6 (a), for nonlinear tests of E0, E5 and E6 (b) and for linear and nonlinear tests of E5(c)

In addition, the result obtained from Fig. 5(c) may well explain the limited field observations. *In-situ* observations of a mode-2 IW generated by a mode-1 IW has been suggested by Liu et al. [25], which shows an extremely large mode-2 IW on 10

April, 1999 in their Fig. 2 and a small mode-2 IW on 24 May, 2009 in their Fig. 8. Furthermore, a remarkable feature in their Fig. 2 is that the amplitude of mode-2 IW even exceeds that of the preceding mode-1 IW, which indicates that the mode-2 IW has been effectively generated in a weak stratification regime in early April as compared to that in late May, which is in agreement with the present findings. Another potential support of the present findings comes from satellite observation. Dong et al. [30] reported the clear signatures of mode-2 IWs closely following mode-1 IWs in April, 2001, which is rarely observed in synthetic aperture radar images. The unusual phenomenon was attributed to an anticyclonic eddy. A close examination of Fig.6 in [30] reveals that the anticyclonic eddy causes the weakening of the pycnocline, a key motivator for the generation of mode-2 IW by mode-1 IWs according to the present study. Meanwhile, satellite observation in the same area after a month does not show mode-2 IWs, probably due to the lack of a weakening pycnocline.

**4.2.4 Effect of front length**

Fig.5 (d) depicts the evolution of $R_2$ for cases E4 (moderate front length), E7 (small front length) and E8 (large front length). The results present no evident differences in the three curves. The occurrence of this phenomenon can be linked with the nearly balanced positive and negative effects of the horizontal stratification on the intermodal interaction. Inspection of Fig.9 reveals that a small front length leads to a stronger linear intermodal interaction, a natural consequence of stronger horizontal stratification in E7, which has $M_0^2 = O(10^{-6}\,\text{s}^{-2})$, approximately an order of magnitude larger than $M_0^2 = O(10^{-7}\,\text{s}^{-2})$ in E4 and E8. On the other hand, the small front length inhibits the nonlinear intermodal interaction to a nearly same degree. Thus, the nearly same trends in Fig.5 (d) result.

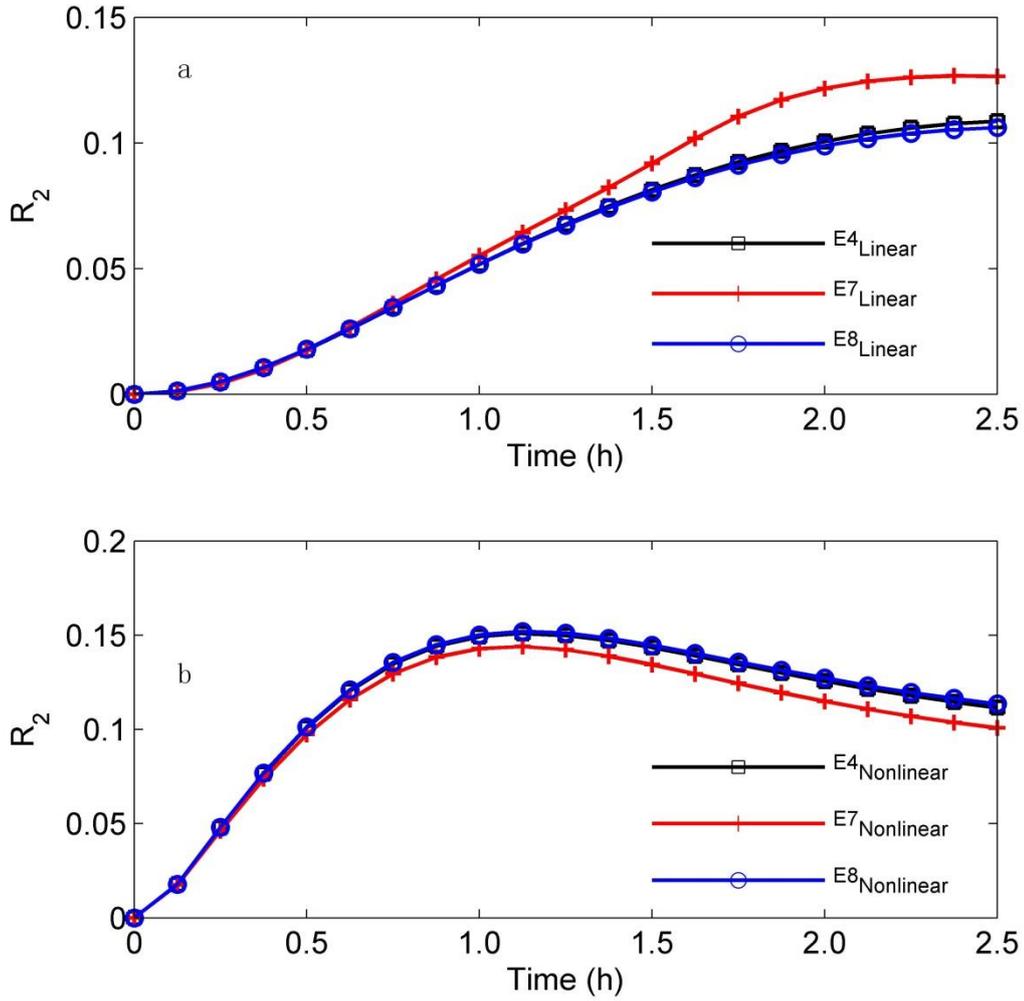

Fig. 9 Comparison of $R_2$ for linear tests of E4, E7 and E8 (a), for nonlinear tests of E4, E7 and E8 (b)

### 4.2.5 Effect of mode-1 wave amplitude

Fig. 10 depicts the impact of amplitude of an initial mode-1 IW. The increase of wave amplitude enhances the nonlinearity of model regime. Specifically, the cases E4, E9 and E10 have the parameter of nonlinearity $\varepsilon = 0.04$, 0.10 and 0.20, respectively. It is seen from Fig.10 that increase of wave amplitude appreciably increases $R_2$ and the mode-2 wave amplitude. This is expected since a larger mode-1 IW brings about stronger linear and nonlinear intermodal interaction. Furthermore, the mode-2 wave amplitude $A_2$ that is located at approximately 17 km increases from 0.21 m through 0.63 m to 1.35 m as the initial mode-1 wave amplitude $A_1$ increases from 2 m through 5.5 m to 10.5 m. Least square fits to the present data yields a relationship between the mode-2 wave amplitude at $t = 2.5h$ and the mode-1 wave amplitude at $t = 0$: $A_2 = 0.14A_1 - 0.087$. The confidence bounds for the coefficients in the linear fit are 95% and the root mean square is 0.047.

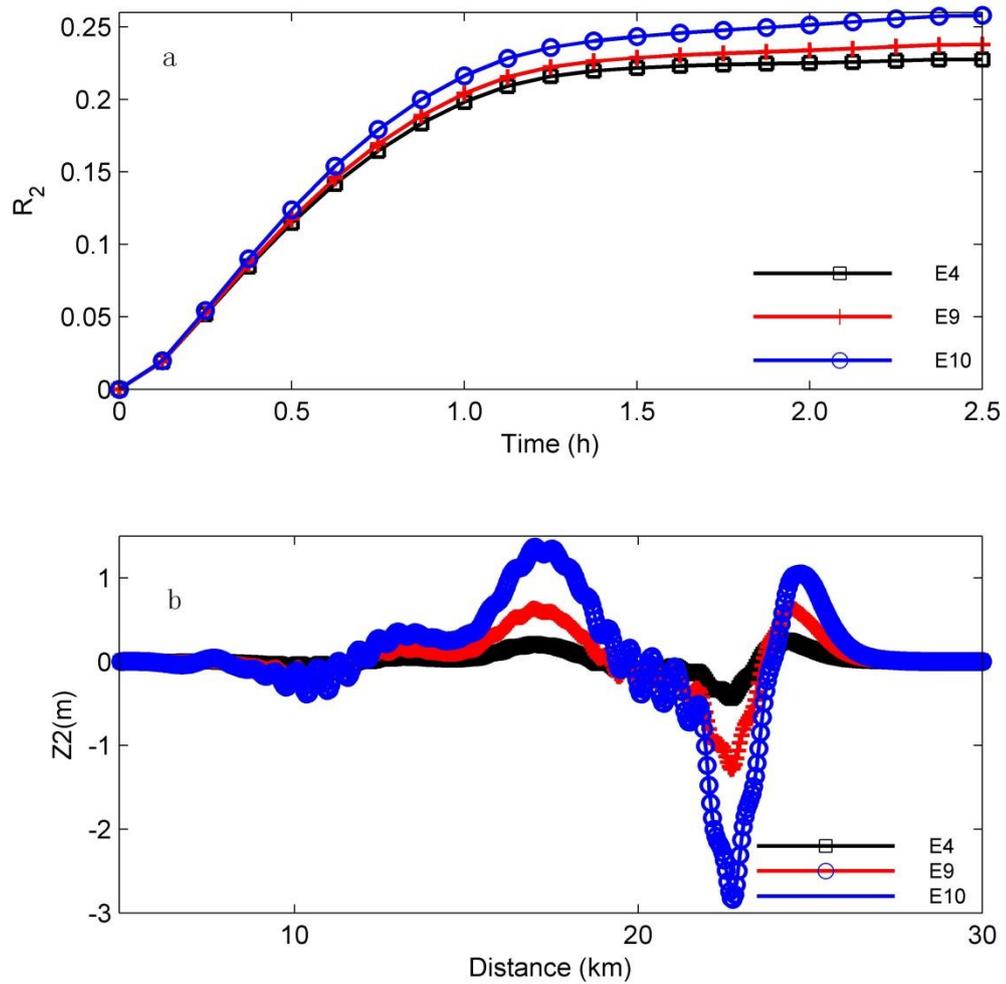

Fig. 10 Comparison of $R_2$ (a) and $Z_2$ (b) for E4, E9 and E10

## 5. Summary

This paper has examined the generation of mode-2 IWs by the evolution of a mode-1 IW in a two-dimensional stratification. A generation model adopting weakly nonlinear and non-hydrostatic approximation was derived based on a multi-modal approach. The generation model accounts for linear and nonlinear intermodal interaction. The nonlinear intermodal interaction results from nonlinear advection and non-hydrostatic effects while the linear intermodal interaction results from horizontal stratification and associated geostrophic current. Evolution of mode-1 and mode-2 IWs in a two-dimensional stratification characterized by an inhomogeneous pycnocline was simulated by numerically solving the generation model. Numerical experiment results were used to expound the details of generation and offer the favorable conditions for the formation of mode-2 IWs.

A mode-2 IW is generated in the wave system when an initial mode-1 IW propagates into a varying stratification regime. The generation of mode-2 IW results from linear and nonlinear intermodal interaction incorporated in the generation model. During the generation process, the mode-2 wave amplitude increases and gradually lags behind the mode-1 IW.

The favorable condition is indicated by larger value of $R_2$ that quantifies the energy amount of mode-2 IWs. The model results suggest that the formation of mode-2 IWs can be largely influenced by pycnocline strength and depth, and less sensitive to pycnocline thickness and front length. A weakening or shoaling pycnocline favors the formation of mode-2 IWs by enhancing linear and nonlinear intermodal interaction, while a thinning pycnocline favors the formation of mode-2 IW mainly by nonlinear intermodal interaction. The nonlinear intermodal interaction is inhibited while the linear intermodal interaction is equivalently strengthened by shortening the front length. Sensitive experiments of nonlinearity shows that increasing an initial mode-1 wave amplitude can appreciably increase the mode-2 wave amplitude by enhancing intermodal interaction.

The previous studies about the generation processes of mode-2 IWs have been restricted to a vertically varying stratification with a distinct topography variation while the present study contributes to expounding the generation process and favorable conditions of mode-2 IWs by a mode-1 IW in a combination of horizontally and vertically varying stratification but with a flat bottom. In practical situations, a large topography variation possibly coexist with a two-dimensional stratification. A fully nonlinear numerical simulation is needed to further explore the generation of mode-2 IW in such a complicated environment.